\documentclass[12pt]{article}
\usepackage{amssymb}
\usepackage{amsmath}
\newcommand {\ov} {\overline}
\newcommand{\ba}{\begin{eqnarray}}
\newcommand{\ea}{\end{eqnarray}}

\newsavebox{\uuunit}
\sbox{\uuunit}
    {\setlength{\unitlength}{0.825em}
     \begin{picture}(0.6,0.7)
        \thinlines
        \put(0,0){\line(1,0){0.5}}
        \put(0.15,0){\line(0,1){0.7}}
        \put(0.35,0){\line(0,1){0.8}}
       \multiput(0.3,0.8)(-0.04,-0.02){12}{\rule{0.5pt}{0.5pt}}
     \end {picture}}

\csname @addtoreset\endcsname{equation}{section}

\newcommand{\IZ}{\mathbb{Z}}
\newcommand{\IP}{\mathbb{P}}
\newcommand{\bp}{\bar{\partial}}
\def\Dslash{\not{\hbox{\kern-4pt $D$}}}
\def\dslash{\not{\hbox{\kern-2pt $\del$}}}
\begin{document}
\begin{titlepage}
\begin{flushright}
SLAC-PUB-11285\\
SU-ITP-05/24\\
hep-th/0507069
\end{flushright}
\vspace{.5cm}
\begin{center}
\baselineskip=16pt
{\bf \LARGE An index for the Dirac operator on D3 branes \\
 \vskip 0.5cm
with background fluxes}\\

\

\

{\large  Eric Bergshoeff$^1$, Renata Kallosh$^{2,3}$, Amir-Kian Kashani-Poor$^{2,4}$, \\
Dmitri Sorokin$^5$,
 Alessandro Tomasiello$^2$
 } \\

\

\

{\small
\centerline{$^1$ Centre for Theoretical Physics, University of Groningen,
Nijenborgh 4,} \centerline{9747 AG Groningen, The Netherlands}\vspace{6pt}
$^2$ Department of Physics, Stanford University,
Stanford, CA 94305-4060, USA.\\ \vspace{6pt}
$^3$ Kyoto University, Yukawa Institute,  Kyoto, 606-8502 JAPAN. \\ \vspace{6pt}
$^4$ SLAC, Stanford University, Stanford, CA 94305-4060, USA.
 }\\ \vspace{6pt}
 \centerline{$^5$ INFN Sezione di Padova ${\&}$ Dipartimento di
Fisica ``Galileo Galilei", } \centerline{Universit\`{a} degli Studi
di Padova, 35131, Padova, Italy}

\end{center}

\

\

\begin{center}
{\bf Abstract}
\end{center}
{\small    We study the problem of instanton generated
superpotentials in Calabi--Yau orientifold compactifications
directly in type IIB string theory. To this end, we derive the Dirac
equation on a Euclidean D3 brane in the presence of background
fluxes. We propose an index which governs whether the generation of
a superpotential in the effective 4d theory by D3 brane instantons
is possible. Applying the formalism to various classes of examples,
including the $K3 \times \frac{T^2}{\IZ_2}$ orientifold, in the
absence and presence of fluxes, we show that our results are
consistent with conclusions attainable via duality from an M-theory
analysis.}

\vspace{2mm} \vfill \hrule width 3.cm
{\footnotesize \noindent e-mails:
E.A.Bergshoeff@rug.nl, \,  kallosh@stanford.edu, \, kashani@slac.stanford.edu, \,  dmitri.sorokin@pd.infn.it, \, tomasiel@stanford.edu }
\end{titlepage}

\section{Introduction}

Stabilization of moduli is an important step on the path towards connecting
M/string theory to effective particle physics and
cosmology in 4d. Many interesting models with stabilization of
all moduli have been discovered over the last few years. We refer
the reader to the most recent paper on this topic \cite{IIA},
which contains a review of the current situation regarding stabilization of moduli in
different versions of string theory and a detailed list of important
references.

To obtain a realistic model, this stabilization should lead to 4d de
Sitter space with a tiny positive cosmological constant to describe
the current acceleration of the universe. Deriving such effective
phenomenological models from M/string theory presents a great
challenge. The technical tools for such an enterprise are rather
limited: in IIA string theory all moduli may be stabilized in anti
de Sitter space using fluxes. The remaining issue is to uplift the
cosmological constant to a positive value; for recent developments on this
issue see \cite{Davidse:2005ef}. The same problem of
uplifting is still to be solved  for heterotic string theory.

On the other hand, in type IIB theory the problem of uplifting has
a reasonable solution. However, stabilization of all moduli in
this theory cannot be achieved at the tree supergravity level,
since fluxes stabilize only complex structure moduli and not
K\"ahler moduli. Thus non-perturbative effects like gaugino
condensation or instanton corrections due to Euclidean D3 branes
wrapped on some 4-cycles of the compactified space should be taken
into account. In \cite{Witten:1996bn} Witten studied instanton
generated superpotentials in M-theory and F-theory
compactifications on CY 4-folds. He derived a necessary constraint
on the divisor the M5 instanton is wrapping in order for the
generation of a superpotential to be possible. The conclusion was
that it must have holomorphic characteristic $\chi=1$. Various aspects of this constraint and its possible modification due to fluxes have been studied in
  \cite{Robbins:2004hx,Gorlich,Kallosh:2005gs,Saulina:2005ve}. The papers \cite{Kallosh:2005gs,Saulina:2005ve} use the Dirac
operator on the M5 brane in the presence of fluxes
\cite{Kallosh:2005yu}; counting the number of
fermionic zero modes in the presence of fluxes and determining the
corresponding flux-dependent index of the Dirac operator leads to
a modification of the $\chi=1$ condition, giving rise to new
possibilities for the stabilization of moduli.

In the interesting cosmological models  \cite{Dasgupta:2002ew}
based on compactification on the $K3\times {T^2\over \IZ_2}$
orientifold with N=2 supersymmetry \cite{Tripathy:2002qw, Andrianopoli:2003jf}, the moduli of the compactification are
distributed in vector and hypermultiplets.  Moduli in vector
multiplets, for example the volume of $K3$,  may be stabilized via
gaugino condensation, since these moduli are the effective gauge
coupling constants for some vector fields. For the moduli in
hypermultiplets, no such terms are available, and one may hope for
instanton generated superpotentials due to Euclidean D3 branes to
achieve the desired stabilization. For example, the volume of the
$ {T^2\over \IZ_2}$ space is a modulus in a hypermultiplet
\cite{Andrianopoli:2003jf} and cannot be stabilized by fluxes
and/or gaugino condensation. Can it be stabilized by instanton
corrections?

>From the results in M-theory \cite{Kallosh:2005gs,Saulina:2005ve}
one may deduce via the connection to F-theory that in type IIB string
theory compactified on $K3\times {T^2\over \IZ_2}$, an instanton
generated superpotential is possible. In the presence of flux that
is of type $(2,1)$ and primitive, the Euclidean D3 branes wrapped
on $K3$ as well as those on $ \IP^1 \times {T^2\over \IZ_2} $
(where $ \IP^1$ is a projective plane within $K3$) can give rise
to superpotentials. Thus, the stabilization of all K\"ahler moduli which were
left unfixed by fluxes becomes possible.

The purpose of this paper is to initiate a systematic study of instanton
corrections to the superpotential directly in type IIB string
theory. This will allow us to draw conclusions on D3 brane instanton
generated superpotentials without having to rely on M-theory/IIB
duality. For earlier results on instanton corrections in type IIB flux
backgrounds see \cite{Tripathy:2005hv}, and also \cite{Gomis:2005wc}, where the problem is considered from the point of view of bosonic zero modes. \vspace{0.5cm}

Our analysis deviates from complete generality in two points
which we would like to point out here:

\begin{itemize}
\item The choice of bulk and worldvolume fluxes:
We constrain our analysis in this paper to the  case that the 2-form
${\cal F}= dA+B_2$ is absent on the Euclidean D3 brane, and we choose the background
flux to be a primitive (2,1)-form. Both restrictions can be removed
in a straightforward way and more general setting of the problem can
be given, however this will lead to more complicated equations. This
we postpone to future studies.

\item Constant axion-dilaton: A generic IIB background including D7 branes
will involve an axion-dilaton field which varies over the internal
manifold. In section \ref{solving}, we consider the simple situation
for which this field is constant.

\end{itemize}

We should also note that in the study of some particular
orientifold examples we encounter the following feature: the
orientifold condition must be compatible with the choice of a gauge
that fixes the $\kappa$-symmetry. Our choice, which is covariant
under duality, is compatible with both the $K3 \times
\frac{T^2}{\IZ_2}$ and
$\frac{T^6}{\IZ_2}$ compactifications, which we discuss in section \ref{example}.
In each new example one should check that the relevant gauge-fixing
condition is compatible with orientifolding.

The paper is organized as follows. We begin in section \ref{sindex}
by proposing an index for the Dirac operator on the D3 brane in the
background of fluxes which, when different from 1, is to rule out
instanton generated superpotentials. In section \ref{brokeandzero},
we comment on the connection between supersymmetries preserved by
the background and fermionic zero modes on the D3 brane. In section
\ref{equation}, we derive the form of the Dirac operator on the D3
brane. In section \ref{solving}, we turn to solving the Dirac
equation. The index proposed in section \ref{sindex} passes its
first test here; it predicts that before orientifolding and turning
on fluxes, instantons do not generate a superpotential. In section
\ref{example} we consider the following orientifolds: $K3 \times
\frac{T^2}{\IZ_2}$, more generally a Fano manifold, and
$\frac{T^6}{\IZ_2}$. We consider brane instantons wrapping several
types of divisors in each of the cases,  compute the corresponding
index and find examples when a superpotential can be
generated. In the discussion section, we explain the significance of adding space-filling D3 branes to our setup and raise the issue of their
mobility.

\section{An index for the Dirac operator on the D3 brane} \label{sindex}
We wish to apply, following Witten \cite{Witten:1996bn}, the
analysis of the zero modes of the Dirac operator on a Euclidean D3
brane to the question of when such instantons can generate
superpotentials in the effective 4d theory. Recall that in
M--theory Witten identifies the charge of the M5 worldvolume
fermions under $U(1)$ transformations in the normal bundle direction with the
R-symmetry of the effective 3d theory. By studying the zero modes of
these fermions, he determines the transformation of the path
integral measure under this normal bundle gauge group. Knowing
that the full theory must be anomaly free (as was established in a
series of subsequent papers \cite{ruben, WittenM, Freed}) allows
him to deduce the effective transformation property of the
instanton action under this normal bundle $U(1)$, from which he can
conclude whether the exponentiated instanton action can play the
role of a superpotential. We wish to perform the same analysis in
the type IIB setup. Note that the anomaly structure of the D3
brane is much simpler than that of the M5 brane, as the
worldvolume theory cannot have a gravitational anomaly. Witten's
analysis however relies on the normal bundle anomaly. As the
worldvolume fermions transform in a complex representation of the
SO(2) normal bundle gauge group, this symmetry is also potentially
anomalous for the type IIB D3 brane. We hence introduce the
following index for the Dirac operator on the D3 brane,
\begin{equation}
\chi_{_{D3}}={1\over 2}( N_+- N_-)\,, \label{index}
\end{equation}
where $N_{\pm}$ is the number of fermionic zero modes with the
$U(1)$ charge $\pm \frac{1}{2}$ in the normal direction.

In the presence of fluxes, the definition of the index
must be modified, since the flux dependence of the Dirac operator
can give rise to zero modes of mixed chirality. Taking into account
that the flux must transform under normal bundle rotations allows us
to assign a definite transformation property to these solutions as
well (which was christened F-chirality in \cite{Kallosh:2005gs}).

Following the reasoning in \cite{Witten:1996bn}, see also \cite{Kallosh:2005gs}, we require $\chi_{_{D3}}=1$ for the generation of a superpotential by instantons to be possible.

\section{ Broken supersymmetry and fermionic zero modes} \label{brokeandzero}

Here we will present a comment on our findings in
\cite{Kallosh:2005yu} and \cite{Kallosh:2005gs} in a form which
generalizes to the D3 brane. In gauge theories, it is well known
that supersymmetries of the background which are broken by gauge
instantons manifest themselves as zero modes of the charged fermions
(one acts on a purely bosonic instanton solution with the
supercharges; those supercharges that are conserved annihilate the
solution, the broken supercharges generate fermionic solutions of
the equations of motion, i.e. in particular of the Dirac equation).
A similar result holds for brane instantons and zero modes of
worldvolume fermions, as we will now argue for M5 branes. The
equations presented in the next section will then demonstrate that
the same arguments go through for D3 branes and are
actually generic for all branes.

The supersymmetries preserved by the background are solutions to
the equation \ba \delta \psi_M \epsilon = 0 \,,  \label{susybg}
\ea where $\delta \psi_M \epsilon$ is the supersymmetry
transformation of the gravitino, i.e. $\delta_\epsilon
\,\psi_M=\delta\psi_M\,\epsilon$ and $M$ is the 11d space--time
index. The supersymmetries preserved by the M5 brane are solutions
to (\ref{susybg}) which satisfy a further constraint
\cite{Becker:1995kb,Bergshoeff:1997kr}. In conventions\footnote{
\cite{Becker:1995kb,Bergshoeff:1997kr} use the opposite sign
conventions.} in which the $\kappa$-symmetry of the brane is given
by $\delta \theta = (1-\Gamma_{M5})\kappa$, this constraint takes
the form
\ba (1 + \Gamma_{M5}) \epsilon = 0 \,. \label{susybrane}
\ea
Supersymmetries satisfying this constraint are preserved,
because the transformation $\delta \theta = \epsilon$ can be
undone by a $\kappa$-symmetry transformation in this case. The
number of supersymmetries broken by the brane are hence
\ba
\#\{\mbox{solutions to (\ref{susybg})} \}- \# \{\mbox{solutions to
(\ref{susybrane})}\} \label{brokensusy}
\,.
\ea
The observation that allows us to
relate this number to the number of zero modes of the worldvolume
fermion is that the linear part of the fermionic equation on the
M5 brane in the presence of background fluxes (but setting the
field strength of the 2-form on the brane to zero) is given by
\begin{equation}
(1-\Gamma_{M5})\,\Gamma^\alpha  \delta \psi_\alpha \, \theta =0\,.
\label{M5Dirac}
\end{equation}
Here, $\alpha$ stands for M5 brane worldvolume indices, bulk
quantities are pulled back onto the brane via $\delta\psi_\alpha =
\delta\psi_M
\partial_\alpha x^M$ and $\Gamma_\alpha = \Gamma_{\underline M}
{e^{\underline M}}_M \partial_\alpha x^M$, the index $M$ is the
curved eleven--dimensional space--time index and $\underline M$ is
the `flat' tangent space index. We see that the pullback of each
solution to (\ref{susybg}) is a solution to the Dirac equation
(\ref{M5Dirac}). Due to $\kappa$-symmetry, not all of these
solutions correspond to dynamical fermions. The most natural
gauge-fixing condition for this symmetry, which we shall call the
duality covariant choice,\footnote{``Duality covariant'' means
that, as one can check, upon the double dimensional reduction the
M5 brane $\kappa$--symmetry projector (and hence the gauge fixing
condition) reduces to that for the D4 brane, which upon a chain of
T--dualities can be transformed into the $\kappa$--symmetry
projectors of other Dp branes \cite{Marolf:2003vf}.} is
\begin{equation}
{1\over 2}(1-\Gamma_{M5})\,\theta = \theta_-  =0\,,\qquad
\theta_{g.f.}=\theta_+\equiv {1\over 2}(1+\Gamma_{M5}) \,\theta\,.
\label{KillingM5}
\end{equation}
Solutions to (\ref{M5Dirac}) which are not dynamical are those which
are pure gauge, i.e. those that are annihilated by the
$\kappa$-symmetry projector,
 \ba (1+\Gamma_{M5})\theta = 0 \,,
\label{gfixing}
 \ea
 and which thus can be removed by the gauge fixing condition,
 like $\theta_-$ in (\ref{KillingM5}).
 The number of dynamical zero modes of the Dirac
equation is hence
 \ba \#\{\mbox{solutions to (\ref{M5Dirac})} \}-
\# \{\mbox{solutions to (\ref{gfixing})}\} \,.
\label{dnm}
 \ea
Comparing (\ref{dnm}) with
(\ref{brokensusy}), the claim that the number of zero modes of the
worldvolume fermion is {\it at least} the number of supersymmetries
of the background broken by the brane follows. But in addition to these zero
modes, the Dirac equation
$\gamma^\alpha \delta \psi_\alpha\,\epsilon=0$ may have more
solutions than those of $\delta\psi_M\,\epsilon=0$. This is also
familiar from gauge instantons: not all fermionic zero modes need to
be related to supersymmetry.

The type of gauge (\ref{KillingM5}) was introduced in
\cite{Kallosh:1995ti,Bergshoeff:1997kr} where the structure of the
Killing spinor of the supergravity background BPS configuration was
used for fixing
$\kappa$-symmetry. This gauge was used in {\cite{Kallosh:2005yu,Kallosh:2005gs}},
where the above conditions were solved in terms of
an irreducible $Spin(1,5)\times Spin(5)$ spinor of the M5 brane
worldvolume, yielding the Dirac equation in gauge fixed form.
 In the general case of
$\kappa$-symmetric superstrings, M2, M5 and Dp branes a similar gauge has been studied from
the geometric perspective of superembedding
\cite{BPSTV,bst,Howe:1997fb,Sorokin:1999jx} ensuring that it is
compatible with any supergravity background  and with the
corresponding gauge fixing of worldvolume diffeomorphisms.

\section{Dirac equation on D3 brane with background fluxes} \label{equation}
 We now focus our attention on the D3 brane. In what
follows we shall put to zero the Dirac--Born--Infeld two--form flux
${\mathcal F}=dA+B_{2}$ on the D3 brane worldvolume. This is admissible for
those compactified solutions of type IIB supergravity for which the
pullback of the NS-NS form
$B_2$ on the D3--brane worldvolume  is pure gauge, i.e. $B_2=-dA|_{D3}$.

Our starting point is the quadratic  Lagrangian for D3 brane
fermions (without the DBI field contribution) derived in
\cite{Marolf:2003vf,Martucci:2005rb} from an M2 brane Lagrangian by
applying a chain of dualities\footnote{A more economic way to get
the Dirac Lagrangian for Dp branes (used for the M2 and M5 brane in
\cite{Kallosh:2005yu}) would be to consider the linearized limit of
the generic Dp brane fermionic equation first obtained in
\cite{bst}.}
\begin{eqnarray}
L^{D3}_f &=& {1\over 2} e^{-\phi}\sqrt{- {\rm det}\, g}\,\bar\theta
(1- \Gamma_{D3}) \big [ \Gamma^\alpha\delta\psi_\alpha
-\delta\lambda\big ]\theta \,. \label{action}
\end{eqnarray}
The equations for unbroken supersymmetry of the background are
\begin{equation}
  \delta\psi_m \, \theta =0 \ , \qquad \delta\lambda \,  \theta=0 \,,
\label{susy}
\end{equation}
(where $m$ stands for a curved 10d index), and the gauge
fixing condition for the action of  $\kappa$--symmetry $\delta_{\kappa} \theta =
(1-\Gamma_{D3})\, \kappa$ reads
\begin{equation}
 (1-\Gamma_{D3})\, \theta=0 \label{g.f.} \,,
\end{equation}
where $\Gamma_{D3}=\sigma_2\gamma_5$, with
$\sigma_2$ acting on a doublet of the Majorana--Weyl spinors
$\theta=(\theta^1,\theta^2)$ and $\gamma_5$ a product of four
ten--dimensional gamma matrices pulled back on the brane.
Comparing to the previous section, we see that the arguments concerning the
number of supersymmetries and fermionic zero modes presented there for the M5 brane apply equally well here.

The detailed form of the action (\ref{action}) has in addition to
the `free' Dirac operator some torsion--type terms  due to
3-fluxes, which we denote by $T_3$ and some terms due to 1--form and
5--form fluxes, which we denote by $T_{(1,5)}$. The action
before gauge-fixing reads
\begin{eqnarray}
L^{D3}_f  &=&  {1\over 2} e^{-\phi}\sqrt{-{\rm
det}\, g}\,\ov\theta \big (1-\sigma_2\gamma_5\big )\bigl
[\Gamma^\alpha\nabla_\alpha + T_3+T_{(1,5)}] \theta
\end{eqnarray}
Here
\begin{eqnarray}\label{t3}
T_3 = {1\over 8}H_{\alpha np}\Gamma^{\alpha
np}\sigma_3 -{1\over 24}H_{mnp}\Gamma^{mnp}\sigma_3+{1\over
8}e^\phi F^\prime_{\alpha np}\Gamma^{\alpha np}\sigma_1 -{1\over
24}e^\phi F^\prime_{mnp}\Gamma^{mnp}\sigma_1
\end{eqnarray}
and
\begin{eqnarray}\label{t15}
T_{(1,5)} = -{1\over 2}\,\Gamma^m\partial_m\,\phi+ {1\over 4}e^\phi
\, F_\alpha\Gamma^\alpha(i\sigma^2) +{1\over{8\cdot
4!}}e^\phi\,F_{\alpha npqr}\Gamma^{\alpha npqr}(i\sigma^2)
\,,
\end{eqnarray}
where $\alpha$ stands for the worldvolume directions of the D3 brane
and the fluxes $H_{3}, F'_{3}, F_5$ and $F_m$ are defined in the
appendix. We basically use the notation and conventions of
\cite{Martucci:2005rb}.

We next impose the gauge fixing condition (\ref{g.f.}), which is
analogous to eq. (\ref{KillingM5}). It can also be given in the
form
\begin{equation}\label{gf11}
\theta_2 = i\gamma_5\theta_1\,,
\end{equation}
which implies that
\begin{equation}\label{gf12}
\theta = \begin{pmatrix} \theta_1\cr i\gamma_5\theta_1 \end{pmatrix} \,,
\hskip 2truecm \ov\theta = \begin{pmatrix}  \theta_1\,, i\theta_1
\gamma_5\end{pmatrix}\,,
\end{equation}
or
\begin{eqnarray}\label{eqs}
{1\over 2} (1+\sigma_2\gamma_5)\theta = \theta \,,&&\hskip 2truecm
{1\over 2} (1-\sigma_2\gamma_5)\theta = 0\,.
\end{eqnarray}
We now push the projection operator ${1\over 2}\big
(1-\sigma_2\gamma_5\big )$ in the Lagrangian through the gamma and
Pauli matrices and let it act on $\theta$ which gives either
$\theta$ back or zero depending on which of the two equations in
(\ref{eqs}) apply, which in turn is determined by whether the operator was commuted past flux with an even or odd number of legs in the D3-brane direction. We
thus find the following expression for the Lagrangian
\begin{eqnarray}\label{fluxlagr}
L_f^{D3} =  \sqrt{-{\rm det}\, g}\,\theta_1\big\{2 e^{-\phi}
\Gamma^\alpha\nabla_\alpha + {1\over 4}{\tilde G}_{\alpha\beta i}
\Gamma^{\alpha \beta i} - {1\over 12} {\tilde G}_{ijk}\Gamma^{ijk}
+{i\over{2\cdot 4!}} \,\gamma_5\,F_{\alpha ijkl}\,\Gamma^{\alpha
ijkl} +{i\over 2}
\,
\nabla_\alpha\tilde\tau\,\Gamma^\alpha\, \big\}\theta_1\nonumber\\
\\
=\sqrt{-{\rm det}\, g}\,\theta_1\big\{2 e^{-\phi}
\Gamma^\alpha\nabla_\alpha + {1\over 4}{\tilde G}_{\alpha\beta i}
\Gamma^{\alpha \beta i} - {1\over 12} {\tilde G}_{ijk}\Gamma^{ijk}
+{i\over{2\cdot 4!}} \,\gamma_5\,F_{\alpha ijkl}\,\Gamma^{\alpha
ijkl}+{i\over 2} \,
\gamma_5\Gamma^\alpha\partial_\alpha
C_{(0)}\,\big\}\theta_1\,,\nonumber
\end{eqnarray}
where $i,j,k,l$ index the 6 directions orthogonal to the D3 brane,
and
\begin{equation}
{\tilde G}_{mnp} \equiv  e^{-\phi}H_{mnp} + i  F_{mnp}'\,\gamma_5\,,
\quad \tilde\tau=C_{(0)}\,\gamma_5+ie^{-\phi}\,.
\end{equation}
  Note that in the kappa--symmetry gauge under consideration the
only $G_3$ flux components which appear in the Dirac Lagrangian are
those which have one or three legs in the directions orthogonal to
the D3 brane, while the contributions of the dilaton $\phi$
derivative vanishes. Note also that because of the self--duality of
$F_5$ and $\Gamma^{m_1\cdots m_5}$ there is only the single term
describing the coupling of the D3--brane fermions to the $F_5$ flux,
which has one leg on the brane. Seemingly differente $F_5$
contributions which appear in the Lagrangian are related by duality
as follows
$$F_{\alpha i_1\cdots i_4}\,\Gamma^{\alpha i_1\cdots
i_4} =-2F_{\alpha
\beta\gamma ij}\,\Gamma^{\alpha\beta\gamma ij}.$$

By duality, the form of the D3 brane Dirac Lagrangian should be
related to the M5 brane Dirac Lagrangian derived in
\cite{Kallosh:2005yu}. Note also that in the chosen duality
covariant gauge the Dirac Lagrangian (\ref{fluxlagr}) has a  simpler
form than in the gauge $\theta_2=0$ imposed in
\cite{Grana:2002tu,Tripathy:2005hv,Martucci:2005rb}. Moreover, as we
shall show in section \ref{example}, it is the duality covariant gauge
fixing condition which is compatible with the $K3 \times \frac{T^2}{\IZ_2}$ orientifold example which is of particular
interest to us for its phenomenological applications.

To study instanton effects, we should now pass from
Minkowski
 to Euclidean signature. In the Dirac Lagrangian (\ref{fluxlagr}) this
 will basically  result in replacing $\sqrt{-{\rm det}\, g}$ with $\sqrt{{\rm det}\, g}$ and
 the Majorana--Weyl spinor $\theta_1$ with a complex Weyl
 spinor, since the Majorana condition is absent in $10d$
 Euclidean space. Note that though the Lagrangian becomes complex,
 the complex conjugate of $\theta_1$ never appears and hence, the
 number of the fermionic degrees of freedom over which the path integral is taken remains the same as in
 space--time of Minkowski signature, i.e. sixteen.

In what follows we shall consider fluxes
\begin{equation}
G_{mnp} \equiv F_{mnp} -\tau H_{mnp}
\end{equation}
that have all legs in the compact 3d complex space,
and we will consider the axion--dilaton $\tau = C_{(0)}
+ie^{-\phi}$
as a constant  that is fixed by fluxes. Under these
assumptions  the Lagrangian (\ref{fluxlagr}) reduces to
\begin{equation}\label{d3fl}
L_f^{D3} = 2 \sqrt{{\rm det}\, g}\,\theta_1\big\{
e^{-\phi}\Gamma^\alpha\nabla_\alpha +{1\over 8}{\tilde G}_{\alpha
\beta i} \Gamma^{\alpha \beta i}\big\}\theta_1\,,
\end{equation}
where the index $i$ is now transverse to the brane and along the internal manifold.
Note that eq. (\ref{d3fl}) coincides, up to an overall
factor, with the D3 brane fermion Lagrangian calculated in the gauge $\theta_2=0$ in
\cite{Tripathy:2005hv}, for the case when all the fluxes except for ${\tilde G}_{\alpha \beta
i}$ are zero (eq. (107) of that paper).
When we consider
the most general
case of non--zero fluxes, the form of
the Dirac Lagrangian in the duality covariant gauge (\ref{fluxlagr}) however
is simpler and differs in some flux terms from the corresponding
Lagrangians in the gauge $\theta_2=0$ (see e.g. eqs. (49), (51) and
(56) of \cite{Tripathy:2005hv} and eq. (73) of
\cite{Martucci:2005rb} with the worldvolume flux set to zero). It
would be of interest to understand this difference in detail, for
instance from the perspective of brane dynamics and gauge fixing
worldvolume diffeomorphisms which should be  compatible with the
$\kappa$--symmetry gauge choice.

\section{Determining the index} \label{solving}
We will perform the calculation of the index in two steps. We first solve the Dirac equation on the covering space of the orientifold and determine the number of zero modes with appropriate sign. We then impose the orientifold condition, which removes some of the modes from the spectrum.

\subsection{Solving the Dirac equation}
The argument in section \ref{sindex} regarding the
possibility of generating a superpotential relies on the
background having
${\cal N}=1$ unbroken supersymmetry in the effective 4d theory.
This means that we should consider only a $G_3$ flux which is (2,1)
and primitive, as shown in \cite{Grana:2000jj}.

Let us review the setup. The D3 brane is wrapped on a 4-cycle $C$
inside the compactified 3-fold/orientifold. The six real dimensions
of the internal complex 3--fold include the directions tangent to
the D3,
$a,b, \ov a, \ov b, $ and normal to it, $z, \ov z$. Rotations in
$z, \ov z$ form the $SO(2)\sim U(1)$ symmetry whose anomaly is
under consideration.

The worldvolume spinor has 10d chirality, so the zero modes will
be of the form $\epsilon_4^+ \otimes \epsilon_6^+$ and
$\epsilon_4^- \otimes \epsilon_6^-$, with the superscripts
indicating the 4d and 6d chirality, respectively. We will omit the
4d factor of the zero mode in the following analysis, with the
understanding that  the internal spinors of even and odd chirality
are tensored with even and odd 4d spinors, respectively, which leads
to a doubling of the number of solutions.

We define the Clifford vacuum as a state $|\Omega\rangle$
 that satisfies the conditions
\begin{equation}
\Gamma^{ z}|\Omega\rangle=0\,, \qquad \Gamma^{ a}|\Omega\rangle=0 \, .
\label{annih}
\end{equation}
All states of fixed chirality in $Spin (10)$ will be divided into
states of positive and negative charge with respect to the normal bundle $U(1)$.
The states with positive chirality are
\begin{equation}
\epsilon_+= \phi |\Omega\rangle + \phi_{\ov a}\Gamma^{\ov a}|
\Omega\rangle\ + \phi_{\ov a \ov b}\Gamma^{\ov a \ov b}|\Omega\rangle \;.
\label{fermions+}
\end{equation}
The states with negative chirality are
\begin{equation}
\epsilon_-= \phi_{\ov z}\Gamma^{\ov z }|\Omega\rangle\ + \phi_{\ov
a \ov z}\Gamma^{\ov a \ov z}|\Omega\rangle\   + \phi_{ \ov z \ov a
\ov b } \Gamma^{\ov z \ov a \ov b }|\Omega\rangle\,.
\label{fermions-}
\end{equation}
Note that
\begin{equation}
{\tilde G}_{(3)}|\Omega\rangle = iG_{(3)}|\Omega\rangle\,,\hskip
2truecm {\tilde G}_{(3)}\Gamma^{\bar a}|\Omega\rangle = -i\bar G_{(3)}
\Gamma^{\bar a}|\Omega\rangle\,.
\end{equation}

We want to consider the effect of a primitive (2,1) 3-form
$G_3$. Let us here consider the simplest case when the value
of the axion-dilaton field is fixed at a constant due to fluxes (note that in a generic
F-theory background, the axion-dilaton field will vary over the
internal manifold). The Dirac equations, obtained from (\ref{d3fl})
and rewritten in terms of (\ref{fermions+}) and
(\ref{fermions-}) are
\begin{equation}\label{1}
\partial_{\ov a}\phi + 4 g^{\ov b c}\partial_c \phi_{\ov b\ov a}
 = 0\,,
\end{equation}
\begin{equation}\label{2}
g^{\ov b a}\partial_a\phi_{\ov b}  = 0\,,
\end{equation}
\begin{equation}\label{3}
\partial_{[\ov a}\phi_{\ov b ]}
=0 \,,
\end{equation}
and
\begin{equation}
\partial^A_{\ov a}\phi_{\ov z} + 4 g^{\ov b c}
\partial^A_c \phi_{\ov b \ov a\ov z}
-i 2 \bar G_{\ov a\ov z b}\phi^{b}=0 \,, \label{flux1}
\end{equation}
\begin{equation}
g^{\ov a b} \partial^A_b\phi_{\ov a\ov z} + 4 i  G_{ab\ov z} \phi^{ab}
=0 \,, \label{flux2}
\end{equation}
\begin{equation}\label{4}
\partial^A_{[\ov a}\phi_{\ov b]\ov z} =0\,.
\end{equation}
On forms which also have a $_{\bar z}$ index, as in \cite{Kallosh:2005gs},
we have a covariant derivative
$\partial^A \equiv \partial + A$ rather than the straight derivative; $A$
is a connection on the line bundle $N=K$. It plays no explicit role in the
following and we will leave it implicit from now on.
The analysis now proceeds exactly as in \cite{Kallosh:2005gs}. First, in
the absence of the flux, by acting with
$\bp\equiv (dz^{\bar a} \wedge)\bp_{\bar a}$ and
$\bp^{\dagger}\equiv (g^{a\bar b}\,\iota_{\partial_{\bar b}})
\partial_a$\footnote{Here $\iota_{\partial_{\bar b}}$ is the contraction with
the vector $\partial_{\bar b}$: it acts as $\iota_{\partial_{\bar
b}} d\bar z^{\bar a_1} \wedge\ldots d\bar z^{\bar a_p} = p\,
\delta_{\bar b}{}^{[\bar a_1} d\bar z^{\bar a_2} \wedge \ldots
\wedge d\bar z^{\bar a_p]}$. The covariant derivative can be replaced
with straight derivatives here because
for a K\"ahler manifold the Levi--Civita connection is non--vanishing only
when all the coefficients are of the same type, as in
$\Gamma^{\bar a}_{\bar b \bar c}$. From $\bp$, it drops because of
antisymmetrization.} on the equations, we see that all
forms must be harmonic. A simple application of Serre duality on the
set (\ref{fermions-}) of negative $U(1)$ charge zero modes shows
their number to be equal to that of positive $U(1)$ charge zero
modes.  We conclude that in the absence of fluxes and if the
orientifold projection is not imposed, the index $\chi_{D3}$
vanishes for brane instantons wrapping 4--cycles in Calabi--Yau
manifolds. This is consistent with the fact that one does not expect
the
${\cal N}=2$ supersymmetry (8 supercharges) of type IIB CY
compactifications to be broken by instanton effects. Note that
this differs from the CY 4-fold compactifications of M-theory,
which can experience superpotential generation due to instantons
in agreement with the fact that these compactifications only
preserve 4 supercharges, i.e. ${\cal N}=1$ supersymmetry (from the
perspective of the effective $4d$ theory).

In the presence of flux, by acting with the harmonic projector
${\cal H}$ (introduced in \cite{Kallosh:2005gs}) on the two flux
dependent equations (\ref{flux1}) and (\ref{flux2}), we conclude
that they have solutions only if the following conditions are
satisfied
\begin{equation}
  \label{eq:harm4}
  {\cal H}  (\bar G_{\ov a\ov z b}\phi^{b})=0\,,
\end{equation}
\begin{equation}
  \label{eq:harm5}
  {\cal H} (G_{ab\ov z}
\phi^{ab})=0\, .
\end{equation}
If solutions for $\phi^b$ or $\phi^{ab}$ exist, then the relation
\begin{equation}
  \label{eq:hdec}
  (1 -{\cal H})(\omega) =
\bar\partial ( \bar\partial^\dagger G \omega)
+ \bar\partial^\dagger (\bar\partial G\omega)\
\end{equation}
with $\omega = 2i \bar G_{\ov a\ov z b}\phi^{b}{}_{\ov m}$
or $\omega = -4 i  G_{ab\ov z} \phi^{ab}$ and $G$ being the Green operator,
can be used to solve, respectively, the equations (\ref{flux1})
and (\ref{flux2}). If we denote solutions to the Dirac
equation by the tuple $(\phi, \phi_{\bar{a}\bar{b}},
\phi_{\bar{a}\bar{z}},\phi_{\bar a}, \phi_{\bar{z}}, \phi_{\bar{a}
\bar{b} \bar{z}})^T$, then the space of solutions is spanned by the
set
\begin{equation}
  \label{eq:recap}
 \left\{ \begin{array}{c}\vspace{.2cm}
\left(
\begin{matrix}
\phi^{harm}\\
0\\
0 \\
0 \\
0\\
0\\
\end{matrix}
\right)  ,
\;\;
\left(
\begin{matrix}
0\\
0\\
\phi_{\ov a \ov z}^{harm} \\
0 \\
0 \\
0 \\
\end{matrix}
\right), \;\; \left(
\begin{matrix}
0\\
0\\
0 \\
0\\
\phi^{harm}_{\ov z}\\
0\\
\end{matrix}
\right), \;\;
\left(
\begin{matrix}
0\\
0\\
0\\
0\\
0\\
\phi^{harm}_{\ov a \ov b \ov z}\\
\end{matrix}
\right)
,\; \;\left(
\begin{matrix}
0\\
0\\
0\\
\tilde{\phi}_{\ov a}\\
g^{\ov a b}\partial_b (G \omega_{\ov a \ov z})\\
\frac{1}{4}\bp_{[\ov a} (G \omega_{\ov b] \ov z})\\
\end{matrix}
\right)
,\;\;
\left(
\begin{matrix}
0\\
\tilde{\phi}_{\ov a \ov b}\\
\bp_{\ov a}( G \omega_{\ov z})\\
0\\
0\\
0\\
\end{matrix}
\right)  \\
 \end{array}  \right\} \,,
\end{equation}
where $\omega_{\ov a \ov z} =  2i \bar G_{\ov a\ov z
b}\tilde{\phi}^{b}$ and $\omega_{\ov z} = -4 i  G_{ab\ov z}
\tilde{\phi}^{ab}$. $\tilde{\phi}_{\ov b}$ and $\tilde{\phi}_{\ov
a \ov b}$ are harmonic forms which in addition satisfy
(\ref{eq:harm4}) and (\ref{eq:harm5}) respectively. The F-chirality weight of these zero modes, in the order presented in equation (\ref{eq:recap}), is $(+---++)$.

\subsection{Orientifolding and the choice of gauge for fixing $\kappa$-symmetry} \label{orientifoldinggauge}
To obtain the Dirac equation in gauge fixed form in section
\ref{equation}, we imposed the gauge (\ref{g.f.}). Before imposing the orientifold projection, we must ensure that this choice of gauge is compatible with the constraint on physical modes which follows from the orientifold projection.

The orientifold action in general is given by (see e.g. \cite{Dabholkar:1997zd})
\begin{equation}
 {\cal O}= (-1)^{F_L} \Omega_P \sigma^* \,,\qquad {\cal O}^2=1 \,,
\label{orient}
\end{equation}
where $\Omega_P$ is worldsheet parity and $\sigma$ is an involution
on the CY manifold. On fermions, $(-1)^{F_L} \Omega_P$ acts by
exchanging the $R-NS$ and $NS-R$ sector, and in addition multiplying
the
$NS-R$ sector by $-1$. Acting on the fermion doublet introduced
above, this operator is represented by $-i \sigma_2$, with
$\sigma_i$ being the usual Pauli matrices. It is useful to compare it with
\cite{Bergshoeff:1999bx} where the transformations of all the fields
are given. Namely, the operation
$(-1)^{F_L} \Omega_P$ in (\ref{orient}) consists of a combination
of two symmetries in \cite{Bergshoeff:1999bx}, one with
$\pm \sigma_3$ acting on fermions and another one with $\pm \sigma_1$,
thus giving
$-\sigma_3 \times \sigma_1= -i\sigma_2$. We should distinguish between two types of brane instanton configurations. If
the worldvolume of the brane is not left invariant by $\sigma$, this action
sends fermions on the brane to fermions on its mirror on the other
side of the orientifold plane. In
this case, the orientifold action is taken into account simply by
disregarding the mirror brane. If, on the contrary, the
worldvolume is sent to itself by $\sigma$, we have an action
$\sigma^*$ on the worldvolume fermions.
This translates into a constraint satisfied by the modes that are
not projected out by the orientifold action, of the general form
 \ba
(1- \Gamma_{\mathcal O}) \,\theta =0  \,.
 \ea
  Now, we need to fix a
gauge for the $\kappa$--symmetry that is compatible with the
orientifolding action. Fixing $\kappa$--symmetry gives rise to a
constraint,
 \ba (1- \Gamma_{\kappa}),\theta =0  \,,
  \ea
   with
$\Gamma_{\kappa}$ depending on the choice of gauge. These two
constraints can be simultaneously imposed if the projectors commute,
i.e. \ba [\Gamma_{\mathcal O}, \Gamma_{\kappa}] = 0 \,. \ea

In the following section, we will study a number of examples. In all
cases, we will find that the duality covariant gauge chosen in
(\ref{g.f.}) is compatible with the orientifolding.

\section{Examples} \label{example}
We now determine the index in some examples which have an
M--theory lift, and verify that both the M--theory and the IIB
analysis yield the same conclusion with regard to the possibility
of generating a superpotential by brane instantons.

\subsection{The $K3\times T^2/\IZ_2$ orientifold}
\label{k3t2} The $K3 \times \frac{T^2}{\IZ_2}$ orientifold can be
obtained as a limit of F-theory on $K3 \times K3$. It is possible to
turn on a flux that stabilizes the 16 D7 branes which are required
for tadpole cancellation on top of the orientifold planes
\cite{Gorlich}. We will assume that this choice has been made, so
that our analysis in the previous section, which was performed with
a  constant axion--dilaton, is applicable.

On the covering space $K3 \times T^2$
of the orientifold, we
will use $1, \ldots, 6$ to denote real internal directions and $1,
\ov 1, \ldots, 3, \ov 3$ to denote complex internal directions, where the first 4 coordinates are in the $K3$ direction, and
the last 2 are in  $T^2$. The involution $\sigma$
in our example acts on both directions $(x^5,x^6)$ of the $T^2$ as
reflection, i.e. $\sigma:(x^5, x^6)\rightarrow (-x^5,-x^6)$.

We will consider three types of D3 brane instanton configurations: on top of an
O7 plane, parallel to it, or intersecting it along one complex
dimension.
\begin{itemize}
\item{
We first consider the case when the D3 brane is on top of an O7 plane, i.e. is wrapping the
$K3$, and is at a fixed point of the
orientifold action in the $T^2$ directions. In terms of the local coordinates introduced above, this implies that the world
volume directions
$a,b$ coincide with the $1,2$ direction, while the normal direction
$z$ is in the $3$ direction.

Let
us  look at the action of $\sigma$ on the fermions. The reflection in the 5 and 6 direction is represented on spinors by $\sigma^* \epsilon(x)= \gamma^{56} \epsilon(\sigma(x))$. But,
due to the results of the previous section, we are only interested
in harmonic forms. In the current example, it so happens that the
coefficients are always even under
$\sigma$, hence we can retain only the multiplication by
$\gamma^{56}$. Since
we are mapping fermions to forms on the brane via
 \ba
\label{iso} \phi_{\bar{a}_1\ldots \bar{a}_n}
d\bar{z}^{\bar{a}_1}\wedge \ldots \wedge d\bar{z}^{\bar{a}_n}
\leftrightarrow \phi_{\bar{a}_1\ldots \bar{a}_n} \gamma^{\bar{a}_1}
\cdots \gamma^{\bar{a}_n} |\Omega \rangle \,,
\ea
we can also read off the action of $\sigma^*$ on fermions by
the action of the pullback on forms: for branes wrapping the $K3$ and
coincident with one of the orientifold planes, we see that
$\sigma^*$ acts with a sign on sections of the normal bundle. This of course coincides with the action given by multiplication of spinors by $\gamma^{56}$.

As explained in subsection (\ref{orientifoldinggauge}), we need to fix a gauge for the $\kappa$ symmetry that is compatible
with the orientifolding action. We first check that the simple gauge
$\theta_2=0$ is not compatible with orientifolding in our model.
The gauge-fixing condition $\theta_2 = 0 $ can also be written in
the form
\begin{equation}
(1-\sigma_3)\theta = 0 \,.
\end{equation}
The orientifold projection would require that
\begin{equation}
(1 -i\sigma_2 \sigma^*)\theta=0 \,.
\label{orient1}
\end{equation}
Thus we have two projectors $(1-\Gamma_\kappa)$ and $(1-\Gamma_{O7})$, with $\Gamma_\kappa=\sigma_3$ and $\Gamma_{O7}=i\sigma_2 \sigma^*$, where $\sigma^*$ commutes with $\sigma_2$. They are anticommuting, and therefore incompatible.

Next, we consider the  gauge-fixing condition for the spinor on
the brane which we used in Section 3 to derive the Dirac operator
(see eqs. (\ref{g.f.}), (\ref{gf11})--(\ref{eqs})), namely
\begin{equation}
(1-\sigma_2 \gamma^5)\theta=0 \,,
\label{Killing}
\end{equation}
where $\gamma^5$ is the product of 4 gamma-matrices on the brane. In the case of the brane wrapping the K3 that we are considering, these directions are $1234$. The gauge fixing condition is hence explicitly given by
\begin{equation}
(1-\sigma_2 \gamma^{1234})\theta=0 \label{Killing1}  \,.
\end{equation}
Now $\Gamma_{\kappa}=\sigma_2 \gamma^{1234}$ and $\Gamma_{O7}=i\sigma_2 \sigma^*$,
where $[\sigma^*,\gamma^{1234}]=[\gamma^{56},\gamma^{1234}]=0$.
 These two projectors commute.
Hence, this gauge is consistent with orientifolding.

To perform our zero mode count, it is convenient to rewrite $\Gamma_{O7}$ and $\Gamma_{\kappa}$ in terms of local complex coordinates,
\begin{equation}
   (1- \Gamma_{\kappa})\theta = (1-\sigma_2 \gamma^{1\ov 1 2\ov 2})\theta=0 \,,
\label{gf1}
\end{equation}
\begin{equation}
   (1- \Gamma_{O7})\theta = (1-\sigma_2 \gamma^{3\ov 3})\theta=0 \,.
\label{or1}
\end{equation}
Together these conditions imply that
\begin{equation}
  (1- \gamma^{1\ov 1 2\ov 2 3\ov 3})\theta=0 \,,
\label{comb}
\end{equation}
which means that both $\theta_1$ and $\theta_2$
are positive chirality spinors in the 6-dimensional
compact space and due to chirality in 10d they are also chiral in $R^4$.
The zero modes with positive $U(1)$
charge, that survive this projection are the
 $2 h^{0,0}$ modes $\phi |\Omega\rangle$, and the $2 h^{0,2}$
 modes $\phi_{\ov a \ov b}\Gamma^{\ov a \ov
b}|\Omega\rangle\ $ . The modes with negative
$U(1)$ charge are the $2 h^{0,1}$ modes
$\phi_{\ov a \ov z}\Gamma^{\ov a \ov z}|\Omega\rangle\ $.
Since the D3 brane is wrapping K3 in this example, which has
$h^{0,1}=0$, the latter modes are absent. Thus without flux contribution
the result for the
index is
\begin{equation}
{1\over 2} (N_+- N_- )= (h^{0,0}+h^{0,2})=2  \,.
\label{K3}
\end{equation}
This is in agreement with the index derived in the M-theory lift of this setup. In M-theory we are using the divisor
$D$ which is a product of $K3_1$ and a $\IP^1$ in a singular
elliptical fiber of $K3_2$. The holomorphic characteristic of this divisor is $\chi_{_{D}}=
\chi_{_{K3\times \IP^1}}= \chi_{_{K3}}  =2$.
This hence excludes the possibility of a flux induced superpotential.

We now add a three--form flux which preserves half of the supercharges and
 fixes the complex moduli in such a way that the D7s are on top of the O7s.
This has been shown to be possible in \cite{Gorlich} in the M--theory
picture. In the IIB picture, at this point in the moduli space
the flux has the form
\begin{equation}
  G_3= c \Omega\wedge d\ov z \ ,
\label{fluxD3}
\end{equation}
where  $c$ is a constant and $\Omega$ is the holomorphic 2-form on $K3$
at this point.
\footnote{This choice of flux also stabilizes
the axion-dilaton and other moduli in this
model\cite{Tripathy:2002qw,Andrianopoli:2003jf}.}. We can now see that the
condition
\begin{equation}
  \label{cut2form}
  {\cal H} (G_{a b  \ov z }\phi^{ab})=0
\end{equation}
forces $\phi_{\ov a \ov b} = 0$, for the following reason. The form $\phi^{ab}$ on K3 is
proportional to $ \ov \Omega$. Thus we have to contract the flux
with $ \ov \Omega$ and we are left with just a harmonic form, so
${\cal H}$ acts trivially on it. This cuts 2$h^{0,2}$ zero modes from
the spectrum, changing the index to
\begin{equation}
  \chi_{_{D3}}(G) = h^{0,0}=1 \,.
\end{equation}
Again, this result reproduces the M-theory analysis in the
presence of a primitive (2,2) flux, and allows us to conclude that
instanton corrections to the superpotential for the modulus which
contains the volume of $K3$ are now possible.}
\item{
When the brane is not on top of one of the orientifold planes but
only parallel, there are a few changes. First of all, even though
the gauge--fixing condition is still (\ref{gf1}), the orientifold
action can be disregarded, since it relates fermions on the brane to
fermions on a mirror brane. Hence, in the absence of
fluxes no modes are cut, and we are left with $2\times (h^{0,0}
+ h^{0,1} + h^{0,2})$ modes both with positive and negative sign.
This gives the index
$\chi_{D3}=0$. This is again consistent  with the M--theory dual:
the elliptic fibration reduces on the D3 brane to a product, and the
M5 dual is thus
$K3\times T^2$, which has vanishing holomorphic characteristic.
Finally, when we add the flux, the relevant equation is again
(\ref{cut2form}), since in this case there are no $\phi_{\bar a}$,
as $h^{0,1}(K3)=0$; as above, the flux is covariantly constant, so
the harmonic projector
${\cal H}$ acts as the identity and all the
$\phi_{\bar a \bar b }$ are cut. However, this time we still have
 the modes $\phi_{\bar z}$ and $\phi_{\bar a \bar b \bar z}$.
The index then becomes
\begin{equation}
  \frac12 (N_+ - N_-) = (h^{0,0}- h^{0,0} - h^{0,2}) = -1
\end{equation}
which is not compatible with superpotential generation.}
\item{
We finally consider a D3 brane wrapping a $\IP^1$ in the K3 and the
$T^2$. In this case, the worldvolume of the D3 brane intersects an O7 locus along one complex dimension. The worldvolume directions $a,b$ are now along $2,3$, while
the normal direction $z$ is along $1$.

By (\ref{iso}), the action of $\sigma^*$ on spinors living on branes wrapping $T^2$ is governed by the action of $\sigma^*$ on $H(T^2)$.
Clearly, elements of $H^0(T^2)$ and $H^2(T^2)$ are even under this action, while elements of $H^1(T^2)$ are odd, thus reproducing once again that $\sigma^*$ acts on spinors via multiplication by
$\gamma^{56}$. The conclusion reached above that the $\kappa$--symmetry gauge $\theta_2=0$ is not compatible with orientifolding hence applies here as well. The duality covariant gauge for the present brane instanton configuration reads
\begin{equation}
(1-\sigma_2 \gamma^{1256})\theta=0 \label{Killing2} \,.
\end{equation}
With $\Gamma_\kappa=\sigma_2 \gamma^{1256}$ and $\Gamma_{O7}=i\sigma_2
\sigma^*$ and
$[\sigma^*,\gamma^{1256}]=[\gamma^{56},\gamma^{1256}]=0$, we conclude that the two
projectors commute.
We see that the duality covariant gauge is compatible with orientifolding for this configuration as well.

The gauge-fixing condition on the brane here reads
\begin{equation}
  (1-\sigma_2 \gamma^{2\ov 2 3\ov 3})\,\theta=0\,.
\label{gf2}
\end{equation}
The orientifold condition requires
\begin{equation}
  (1-\sigma_2 \gamma^{3\ov 3})\,\theta=0\,.
\label{or2}
\end{equation}
Together these conditions imply that
\begin{equation}
  (1- \gamma^{ 2\ov 2 })\,\theta=0 \;.
\label{comb2}
\end{equation}
The zero modes that survive this projection are the $2h^{0,0}(\IP^1
\times T^2)=2$ states from $\phi |\Omega\rangle$ and the
$2h^{0,1}(T^2)=2$ states from $\phi_{\ov 3} |\Omega\rangle$ of a
positive $U(1)$ charge, as well as $2 h^{0,2}(\IP^1 \times T^2)=0$
modes from $\phi_{\ov z} |\Omega\rangle$ and $2 h^{0,1}(\IP^1)=0$
modes from $\phi_{\ov 3 \ov z} |\Omega\rangle$ of a negative $U(1)$
charge, where we have used Serre duality to count the number of the
negative $U(1)$ charge modes.
 As a result, in the absence of fluxes the index is
\begin{equation}
{1\over 2}( N_+- N_-) = 2 \,.
\label{P1*P1}
\end{equation}

Note that the M-theory lift of this divisor is the same as in the previous example,
$D=K3\times \IP^1$, and $\chi_D = 2$.

In the presence of flux, the 1-form $\phi_{\ov 3}$ must satisfy the additional constraint (\ref{eq:harm4})
\begin{equation}
  \label{eq:harm44}
  {\cal H}  (\bar G_{\ov 1\ov 2 3}\phi^{3})=0\ .
\end{equation}
Since the argument of the projector is a multiple of the $(0,2)$
form of the K3, the only solution of this constraint is $\phi_{\ov
3} = 0$. Thus, when the flux is turned on only the two zero modes
from $\phi |\Omega \rangle$ survive  yielding the index $\chi=1$.
As in the M-theory analysis, the conclusion is that in the
presence of the flux the instantons can generate a superpotential
also in this case.}
\end{itemize}

\subsection{D3-branes on general Fano manifolds}

We can now generalize the $K3\times \frac{T^2}{\IZ_2}$ example, by considering the base manifold of the elliptically fibered F-theory fourfold to be an arbitrary Fano manifold.\footnote{By definition, Fano manifolds have positive anticanonical bundle. This condition will come into play below equation (\ref{whyfano}).}

A general elliptic fibration is described by the usual Weierstrass
equation describing an elliptic curve, $y^2=x^3 + f x + g$, but where
$f$ and $g$ are let vary over a base manifold $B$. For consistency of
the equation, it turns out that $f$ and $g$ must be sections of ${\cal L}^4$
and ${\cal L}^6$ respectively, where ${\cal L}$ is a line bundle over $B$.
If we want the total space of the fibration to be a Calabi--Yau, we have to
take ${\cal L}=K^{-1}$, the inverse of the canonical line bundle of $B$.
The axion--dilaton is, in general, determined by $f$ and $g$ through $j(\tau)=
\frac{24}{1 + (3^3 g^2)/(2^2 f^3)}$, where $j(\tau)$ is a certain known
function. In general, $\tau$ will vary and will be outside perturbative control
over some region of the base; but we can also make it to be constant if we
choose $f$ and $g$ appropriately. Since $\tau$ only depends on the ratio
$g^2/f^3$, if this ratio is constant so is $\tau$. This can be solved by
taking
\begin{equation}
f=a q^2\ , \qquad g=b q^3\ , \label{whyfano}
\end{equation}
with $a$ and $b$ being some constants and
$q$ being a section of $K^{-2}$. If $B$ is a Fano manifold, by definition $K^{-1}$ is positive and hence has holomorphic sections, as does $K^{-2}$. We will call $C$
the zero locus of $q$.

All this is essentially a generalization of the example $B=\frac{T^2}{\IZ_2}=
\IP^1$ considered in \cite{sen1} and (taking the product with a
spectator $K3$) in section \ref{k3t2}. There, $K^{-1}$ has degree
2, so $K^{-2}$
has degree 4. Thus $q$ has 4 zeros.
The monodromy argument of \cite{sen1} tells us that on each of these
4 zeros there is an O7 and four D7s. This is consistent with the fact that
the axion--dilaton is constant, since the tadpole is canceled locally.
The monodromy argument is local and also goes through for a general Fano $B$,
going around the zero locus of $q$. So on this zero locus there is
an O7 with 4 D7s, once again. The manifold $B'$
whose $\IZ_2$ quotient gives $B$ (the generalization of $T^2$ in the $\IP^1$
example above) can be described as in \cite{sen2} as a double (branched)
covering of $B$ by the equation $\xi^2=q$; the $B'$ thus described is also
a Calabi--Yau.

We now want to put a D3 on top of the zero locus of $q$ (or of one
of its components, if it has more). The analysis of the first case
in section \ref{k3t2} goes through, and we repeat it here. First of
all, the gauge fixing and the orientifold projection look locally
around the brane like (\ref{gf1}) and (\ref{or1}). Then we can once
again combine them to get a condition (\ref{comb}). This says that
$\theta_i$ have positive internal chirality. This leaves us with
only the modes $\phi$, $\phi_{\bar a \bar b}$ (to be counted with +
sign) and $\phi_{\bar a \bar z}$, which can be dualized to a
$\tilde\phi_{\bar b}$ (to be counted with a sign -). So in this
case the index is
\begin{equation}
  \chi_{D3}=h^{0,0}- h^{0,1}+ h^{0,2}\ ,
\end{equation}
 which just happens to be the holomorphic $\chi$ of the
zero locus $C$.

Before turning on the flux, we can check that this index coincides
with the one in M--theory. The locus $C$ on which the D3 brane is
sitting is exactly where the elliptic fibration gets singular. The
blow--up of this singularity gives a chain of $\IP^1$'s intersecting
in points. The dual of the D3 in M--theory is then wrapping one of
these
$\IP^1$ along the fibre, and $C$ along the base. The holomorphic
characteristic of this lift is then
\begin{equation}
\chi_{M5}=\chi(\IP^1)\times \chi_{D3}=
\chi_{D3}
\end{equation}
 (since $\chi(\IP^1)=1$).

When we add a flux, we are led again to consider the equations of
section \ref{solving}. The only difference is that now some of the
modes are not there because they have been projected away. We are
now left with the sole condition  ${\cal H}(G_{ab\bar z}
\phi^{ab})=0$, which determines whether or not the mode $\phi_{\bar
a \bar b}$ is lifted. This of course has to be determined case by
case, but we can once again check consistency with the M--theory
picture. Indeed, the M--theory dual of $F$ and $H$ is
$F_{\mathrm{M-theory}}= G\wedge d\bar\zeta$, where $\zeta$ is
the holomorphic coordinate on the fibre, as follows from supersymmetry.
 Using this, the condition
${\cal H}(G_{ab\bar z} \phi^{ab})=0$
is dual in M--theory to the condition
\begin{equation}
{\cal H}(F_{a b \bar c \bar z}\phi^{ab})=0\ ,
\end{equation}
which is just the condition
obtained in \cite{Kallosh:2005gs}.

As a final comment, we can also consider the case in which the D3
brane sits on a locus $D$ which never intersects $C$, generalizing
the second brane instanton configuration of section \ref{k3t2}. In that case, the
orientifold projection only relates the D3 to a mirror D3, and it
does not cut any of the modes. Thus the index is clearly zero. On
the other hand, the elliptic fibration is never degenerating on
$C$, since the D3 never intersects the degenerating locus $C$. So
the dual M5 is a non--singular $T^2$ fibration over $C$, and its
holomorphic characteristic gives $\chi_{M5}= \chi(T^2)\times
\chi_{D3}=0$ (since
$\chi(T^2)=0$). So the two indices agree once again.

One might wonder whether a generalization of the third brane instanton configuration of
section \ref{k3t2} is also possible. One could consider a case in
which the D3 brane locus $D$ has a non--zero intersection with the
O7/D7 locus
$C$. It is certainly possible to compute the holomorphic
characteristic of the dual M5 (for example as in
\cite{Robbins:2004hx}), but a general computation is much less clear
in the type IIB setting. It would be interesting to explore this
class of examples, and compare them with its M--theory dual.

\subsection{The case $T^6/\IZ_2$}
Finally, we will briefly discuss an example with O3 rather than
O7 orientifold planes, and see how it also fits into our analysis. Let us consider an orientifold defined
by
$\Omega_P (-1)^{F_L}\sigma^*$, where $\sigma$ is now the reflection
of {\it all} the coordinates of $T^6$.  This configuration involves 64 O3 planes. We can consider two types of D3 brane instanton configurations: a D3 brane on top of an O3 plane, or away from it. In the first case, the orientifold
projection keeps only modes that satisfy the constraint
\begin{equation}
   (1-i \sigma_2 \gamma^{x^1 y^1 x^2 y^2 x^3 y^3})
\theta\equiv (1-\Gamma_1) \theta =0\ .
 \label{o3}
 \end{equation}
As in the $K3\times \frac{T^2}{\IZ_2}$ example, there would also be
an action of $\sigma$ on the argument of the spinor, but since we
are dealing here with harmonic forms only (for which the
coefficients are all constant in this case, and in particular even),
we can disregard the argument altogether. On the other hand, when the D3 brane is not on top of one of the O3 planes, the worldvolume of the brane is sent to its mirror by $\sigma$, and
the orientifold projection can be disregarded.

Now let's consider the question of choosing a gauge for $\kappa$--symmetry compatible with the constraint (\ref{o3}).
$\theta_2=0$ is not compatible with this constraint: as before, we can write this choice of gauge as $(1-\sigma_3) \theta=0$; since $\Gamma_1$ involves $\sigma_2$, the constraint $\theta_2=0$ clearly does not commute with the orientifold constraint (\ref{o3}).

 Next, we consider the  duality covariant gauge-fixing condition for
 the spinor on the brane of the form
 \begin{equation}
 (1-\sigma_2 \gamma^5)\theta \equiv (1-\Gamma_{O3})\theta =0\,.
 \label{Killing3}
 \end{equation}
Recall that $\gamma^5$ is the product of the 4 gamma-matrices on the brane. Since $[\Gamma_\kappa,\Gamma_{O3}]=0$, this gauge is once again the one that is consistent with orientifolding.


Together, (\ref{Killing3}) and (\ref{o3}) give
\begin{equation}
   (1- \gamma^{ z \ov z})\theta =0  \,.
 \label{orient2}
 \end{equation}
This cuts all states in $\epsilon_-$. The total number
 of zero modes is now only $2\times (h^{0,0} + h^{0,1} + h^{0,2}) = 2 ( 1 + 2 + 1) =8$
 and the index is equal to 4.

We can compare this 4 with the M--theory dual $T^8/\IZ_2$ \cite{sen3}.
The dual M5 wraps a $T^2/\IZ_2$ locus. The holomorphic characteristic of
this orbifold is simply $h^{0,0} + h^{2,0}=4$ (since odd cohomology is
projected out by the $\IZ_2$ action).

Let us now add the flux. For the model of $\frac{T^6}{\IZ_2}$ the
flux which preserves ${\cal N}=1$ supersymmetry (but not larger) is
of the form
 \cite{Kachru:2002he}
\begin{equation}
   G_3\sim dz^1\wedge dz^2\wedge d\ov z^3 +dz^2\wedge dz^3\wedge d\ov
 z^1+dz^3\wedge dz^1\wedge d\ov z^2\ ,
 \label{flux}
\end{equation}
where $z^i=x^i + i y^i$.
This is once again covariantly constant, so in the conditions
(\ref{eq:harm4}) and (\ref{eq:harm5}) the contraction of the modes
$\phi_{\bar a \bar b}$ and $\phi_{\bar a}$ with $G$ is automatically
harmonic. This means that $h^{0,1}$ and $h^{0,2}$ are cut. Hence we
are left with
$2h^{0,0}=2$ positive chirality zero modes. The index equals 1, and
we have instanton corrections.

Finally, let us consider the case in which no orientifold plane
meets the brane, a case which has been considered recently in
\cite{Tripathy:2005hv}. The orientifold projector can now be
disregarded. Before introducing  the flux, we have the modes
$\phi, \phi_{\bar a}, \phi_{\bar a \bar b}$ and $\phi_{\bar z},
\phi_{\bar a \bar z}, \phi_{\bar a \bar b\bar z}$, which exactly
cancel each other, leaving the index 0. In the presence of fluxes,
the relevant equations are now (\ref{eq:harm4}) and
(\ref{eq:harm5}). Since the flux (\ref{flux}) is covariantly
constant, as in the cases above the harmonic projector
${\cal H}$ acts as the identity, and all the modes $\phi_{\bar a}$ and
$\phi_{\bar a \bar b}$ are lifted. This leaves us with the index
\begin{equation}
 \chi_{D3}= h^{0,0}- h^{0,0}- h^{0,1} - h^{0,2}= -2-1=-3
\end{equation}
which is incompatible with superpotential generation and agrees with the results of
\cite{Tripathy:2005hv}.

\section{Discussion}

In this paper we have demonstrated  that the coupling of
worldvolume fermions to fluxes opens up the possibility of
having instanton corrections to the superpotential from Euclidean D3
branes in cases where such corrections were ruled out in the absence
of flux. We have introduced an index for the Dirac operator on
the D3 brane,
\begin{equation}
\chi_{_{D3}}={1\over 2}( N_+- N_-)\,, \label{index1}
\end{equation}
where $N_{\pm}$ is the
 number of fermionic zero modes having a $U(1)$ charge $\pm \frac{1}{2}$
 in the normal direction.
 We argue that ${1\over 2}( N_+- N_-)$
 must be equal to 1
 to allow instanton corrections to the superpotential.
 In the presence of fluxes,
 the counting of the fermionic zero modes weighted
 with the $U(1)$ charge may be modified, compared to the cases without fluxes,
 as we have shown in some interesting examples.

In particular, in type IIB string theory compactified on a $K3\times
T^2/\IZ_2$ orientifold, it is possible to have instanton corrections
to the superpotential for the D3 brane wrapped on $K3$ at the
orientifold locus as well as on $\IP^1\times T^2/\IZ_2$ with $\IP^1$
being a curve in the $K3$. This  is in agreement with the
corresponding results in M-theory established in
\cite{Kallosh:2005gs}. In the presence of such contributions to the
superpotential, the stabilization of all moduli in type IIB string
theory compactified on a $K3\times T^2/\IZ_2$ orientifold is within
reach, and in fact has been achieved from the perspective of
M-theory compactified on
$K3\times K3$ in \cite{AK} and in its F--theory dual.

For cosmological applications, it is interesting to add D3 branes extended along the physical non-compact directions to the mix. In the tree--level analysis, the positions of the D7 branes
are generically fixed by the  flux, whereas the positions of the D3
branes remain unfixed moduli as long as no two--form along the D7
worldvolume is introduced. The D3 brane positions in $ T^2/\IZ_2$
belong to vector multiplets and the positions in $K3$ belong to
hypermultiplets. In the model of \cite{Dasgupta:2002ew,Hsu:2003cy}
the distance between the D3 and D7 branes serves as the inflaton.
This model has the attractive feature that instead of having to
fine-tune the inflaton potential, its flatness arises due to a
dynamical mechanism, namely the slightly broken shift symmetry
\cite{Hsu:2003cy,Dasgupta:2002ew} in the inflaton direction due to a
not-self-dual 2-form flux on the D7 branes. In the presence of 2-form
flux on the D7s the cosmological system has 2 stages of evolution, one in
a locally stable de Sitter valley with almost flat directions when
the D3 is far from the D7s,  and the second stage when the absolute minimum
of the potential is reached, when both the D7s and the D3 are stabilized and
the D-flatness condition is restored. In many recent  studies of
flux vacua this feature of cosmological D-term inflation model
has not been appreciated yet and included in the analysis.
Other features of the model include a well understood exit from the
inflation stage with reheating and the value of the tilt of the
spectrum $n_s=0.98$ which is the central value of the current data.

While K\"ahler moduli and D3 brane positions enter the action in different ways\footnote{K\"ahler moduli are given by volumes of various
4-cycles in the $K3\times T^2/\IZ_2$ orientifold, and as such appear
in the classical action of brane instantons. Instanton corrections
to the superpotential hence give rise to a dependence of the form
$\sim e^{-\rho}$, with
$\rho$ indicating a generic K\"ahler modulus. In contrast, the
positions of D3 branes enter more delicately in the theory, e.g.
determining the masses of modes of strings stretching between D3 and
D7 branes.}, one can ask the question whether the latter are also
lifted in the presence of brane instantons.
 For example, dependence on these positions could arise in the determinant
prefactor, as argued in \cite{Ganor} in the absence of fluxes, and
in \cite{Berglund:2005dm}, whose underlying assumption is that the
duality between $K3\times K3$ and type II on a Calabi--Yau is still
valid in the presence of fluxes. In the related context of gaugino
condensation giving rise to superpotentials, \cite{Berg:2004sj}
argue for a dependence on D3 brane positions arising due to
threshold corrections.  More work is required to clarify the status
of D3 brane positions in the context of the model presented in our
paper, with instanton corrections rather than gaugino condensation
giving rise to K\"ahler moduli dependence in the superpotential.

If the mobility of D3 branes in the de Sitter valley when far from the D7s survives instanton corrections, string
cosmology will receive a significant boost. This will constitute the
last step in deriving D--term inflation
\cite{DBH} with its various attractive phenomenological features from string theory.

\bigskip\bigskip

\leftline{\bf Acknowledgments}

\

We are grateful to P. Aspinwall, M. Douglas,
 S. Kachru, A. Linde, D. L\"ust, L. Martucci, F. Marchesano, J. McGreevy,
S. Reffert, S. Stieberger, P. K. Tripathy and S. Trivedi for
useful discussions.
This work was supported by NSF grant 0244728. The work of AK was
also supported by the U.S. Department of Energy under contract
number DE-AC02-76SF00515. The work of E.B. and D. S. was supported
by the EU MRTN-CT-2004-005104 grant `Forces Universe' in which E.B.
is associated to Utrecht University, and by the MIUR contract no.
2003023852 (D.S.).
\
\\
\\
\

\begin{appendix}
\section{Conventions and definitions}

We denote (Lorenzian) worldvolume indices by
$\alpha = 0,1,2,3$ and (Lorenzian) target space indices by
$m=0,1,2,\cdots ,9$. The corresponding flat indices are
underlined. The ten-dimensional chiral operator is
$\Gamma_{11} = \Gamma_{\underline{0\cdots 9}}$ with
$(\Gamma_{11})^2 = 1$.

The matrix $\Gamma_{D3}$ of the kappa--symmetry projector is defined
by
\begin{equation}
\Gamma_{D3} = e^{-a/2}\, \sigma_2\gamma_5 \,e^{+a/2}\,,
\end{equation}
with $a({\cal F})$ some function of ${\cal F}$ \cite{Bergshoeff:1997kr} and
$\gamma_5 = i\gamma_{\underline{0123}}\,,\ \
(\gamma_5)^2 = 1$ \footnote{For the passage from Minkowski to
Euclidean space it is important to remember that in Euclidean space
$\gamma_5$ with $(\gamma_5)^2=1$ is defined as $\gamma_5 =
\gamma_{\underline{1234}}$ and hence is {\it the same as}
$\gamma_5$ in Minkowski space since $\gamma_{\underline 4}=-i\gamma_{\underline 0}$. }. In
the absence of worldvolume flux, i.e.
${\cal F}_{\alpha\beta}=0$, we have $a=0$ and the expression for
$\Gamma_{D3}$ reduces to

\begin{equation}
\Gamma_{D3} = \sigma_2\, \gamma_5\, .
\end{equation}

The gravitino and dilatino supersymmetry rules are given by
$\delta_\epsilon \psi_m= \delta\psi_m \epsilon $ and
$\delta_\epsilon \lambda= \delta\lambda \epsilon $
where $\epsilon$ is a doublet of 10d chiral spinors and
\begin{equation}
\delta\psi_m = \delta\psi_m^{NS} + \delta\psi_m^{RR}\, ,\hskip 1truecm
\delta\lambda = \delta\lambda^{NS} + \delta\lambda^{RR}\,,
\end{equation}
with

\begin{eqnarray}\label{localsusy}
\delta\psi_m^{NS}&=& \nabla_m + {1\over 4\cdot 2!}H_{mnp}\,\tilde\Gamma^{np}
\sigma_3\,,\\
\delta\psi_m^{RR}&=& {1\over 8}e^\phi\big[
-F_n\tilde\Gamma^n(i\sigma_2) +{1\over 3!}{
F}_{npq}'\tilde\Gamma^{npq}\sigma_1
+ {1\over 2\cdot 5!}F_{npqrt}\tilde\Gamma^{npqrt}(i\sigma_2)\big]\Gamma_m\,,\label{localsusy1}\\
\delta\lambda^{NS} &=& {1\over 2}\big(\Gamma^m\partial_m\phi
+ {1\over 2\cdot 3!}H_{mnp}\Gamma^{mnp}\sigma_3\big)\,,\\
\delta\lambda^{RR} &=& -{1\over 2}e^\phi\big[-F_m\Gamma^m(i\sigma_2)+
{1\over 2\cdot 3!}
 F_{mnp}'\Gamma^{mnp}\sigma_1\big]\,,
\end{eqnarray}
where $\nabla_m = \partial_m + {1\over 4}\Omega_m{}^{\underline{np}}
\tilde\Gamma_{\underline{np}}$ is the covariant derivative. The different curvatures
appearing in the supersymmetry rules are defined by $H_{(3)} = dB_{(2)},
F_{(3)} = dC_{(2)}$ and

\begin{equation}
F_{(1)} = dC_{(0)}\,,\hskip 1truecm
{ F}_{(3)}' = F_{(3)} - C_{(0)} H_{(3)}\,,\hskip 1truecm
F_{(5)} = dC_{(4)} + H_{(3)}\wedge C_{(2)}\,.
\end{equation}

Note that in eqs. (\ref{localsusy}) and (\ref{localsusy1}) we put
{\it tilde} on antisymmetric combinations of $16\times 16$
gamma--matrices to indicate that they carry the chiral spinor
indices {\it opposite} to those in eqs. (\ref{t3}) and (\ref{t15}).
This distinction is important
 because e.g. $\tilde\Gamma^{m_1\cdots m_5}$ are {\it
anti--self--dual} while $\Gamma^{m_1\cdots m_5}$ are {\it
self--dual}.

\end{appendix}

\end{document}